\documentclass{evn2004}
\usepackage{txfonts}
\usepackage{graphicx}
\begin{document}
   \title{Densification of the International Celestial Reference Frame:
          Results of EVN+ Observations}

   \author{P. Charlot\inst{1}
          \and
          A. L. Fey\inst{2}
          \and
          C. S. Jacobs\inst{3}
          \and
          C. Ma\inst{4}
          \and
          O. J. Sovers\inst{5}
          \and
          A. Baudry\inst{1}
          }

   \institute{Observatoire de Bordeaux (OASU) -- CNRS/UMR 5804,
              BP 89, 33270 Floirac, France
         \and
             U. S. Naval Observatory, 3450 Massachusetts Avenue NW,
             DC 20392-5420, USA
         \and
             Jet Propulsion Laboratory, California Institute of Technology,
             4800 Oak Grove Drive, Pasadena, CA 91109, USA
         \and
             National Aeronautics and Space Administration, Goddard Space
             Flight Center, Greenbelt, MD 20771, USA
         \and
             Remote Sensing Analysis Systems, 2235 N. Lake Avenue,
             Altadena, CA 91101, USA
             }

   \abstract{
The current realization of the International Celestial Reference
Frame (ICRF) comprises a total of 717~extragalactic radio sources
distributed over the entire sky. An observing program has been
developed to densify the ICRF in the northern sky using the European 
VLBI network (EVN) and other radio telescopes in Spitsbergen, Canada 
and USA. Altogether, 150~new sources selected from the Jodrell 
Bank--VLA Astrometric Survey were observed during three such 
EVN+ experiments conducted in 2000, 2002 and 2003. The sources were 
selected on the basis of their sky location in order to fill the 
``empty'' regions of the frame. A secondary criterion was based on 
source compactness to limit structural effects in the astrometric 
measurements. All 150~new sources have been successfully 
detected and the precision of the estimated coordinates in right 
ascension and declination is better than 1~milliarcsecond (mas) for 
most of them. A comparison with the astrometric positions from
the Very Long baseline Array Calibrator Survey for 129 common 
sources indicates agreement within 2~mas for 80\% of the sources.
}

   \maketitle
%
%________________________________________________________________

\section{Introduction}

The International Celestial Reference Frame (ICRF), the most recent 
realization of the VLBI celestial frame, is currently defined by the 
radio positions of 212~extragalactic sources observed by VLBI between 
August~1979 and July~1995 (Ma et al. \cite{Ma98}). These {\it defining} 
sources, distributed over the entire sky, set the initial direction 
of the ICRF axes and were chosen based on their observing histories 
with the geodetic networks and the accuracy and stability of their 
position estimates. The accuracy of the individual source positions 
is as small as 0.25~milliarcsecond (mas) while the orientation of the
frame is good to the 0.02~mas level. Positions for 294~less-observed 
{\it candidate} sources and 102~{\it other} sources with less-stable 
coordinates were also reported, primarily to densify the frame. 
Continued observations through May~2002 have provided positions for 
an additional 109~new sources and refined coordinates for candidate 
and ``other'' sources (Fey et al.~\cite{Fey04}).

The current ICRF with a total of 717~sources has an average of one
source per $8^{\circ} \times 8^{\circ}$ on the sky. While this
density is sufficient for geodetic applications, it is clearly too
sparse for differential-VLBI applications (spacecraft navigation,
phase-referencing of weak targets), which require reference
calibrators within a-few-degree angular separation, or for linking
other reference frames (e.g. at optical wavelengths) to the ICRF.
Additionally, the frame suffers from a inhomogeneous distribution 
of the sources. For example, the angular distance to the nearest 
ICRF source for any randomly-chosen sky location can be as large as 
$13^{\circ}$ in the northern sky and $15^{\circ}$ in the southern sky 
(Charlot et al. \cite{Cha00}). This non-uniform source distribution 
makes it difficult to assess and control any local deformations in 
the frame. Such deformations might be caused by tropospheric
propagation effects and apparent source motions due to
variable intrinsic structure (see Ma et al. \cite{Ma98}).

This paper reports results of astrometric VLBI observations of
150~new sources to densify the ICRF in the northern sky. These
observations were carried out using the European VLBI Network
(EVN) and additional geodetic antennas that joined the EVN for
this project. The approach used in selecting the new potential
ICRF sources was designed to improve the overall source
distribution of the ICRF. Sources with no or limited extended
emission were preferably selected to guarantee high astrometric
suitably. Sections~2 and~3 below describe the source selection
strategy in further details, the network and observing scheme used
in these EVN+ experiments, and the data analysis. The astrometric 
results that have been obtained are discussed in Sect.~4, including 
a comparison with the Very Long Baseline Array (VLBA) Calibrator 
Survey astrometric positions for 129~common sources.

\section{Strategy for Selecting New ICRF Sources}

\begin{figure*}
\centering
\includegraphics[angle=0,origin=br,scale=0.50]{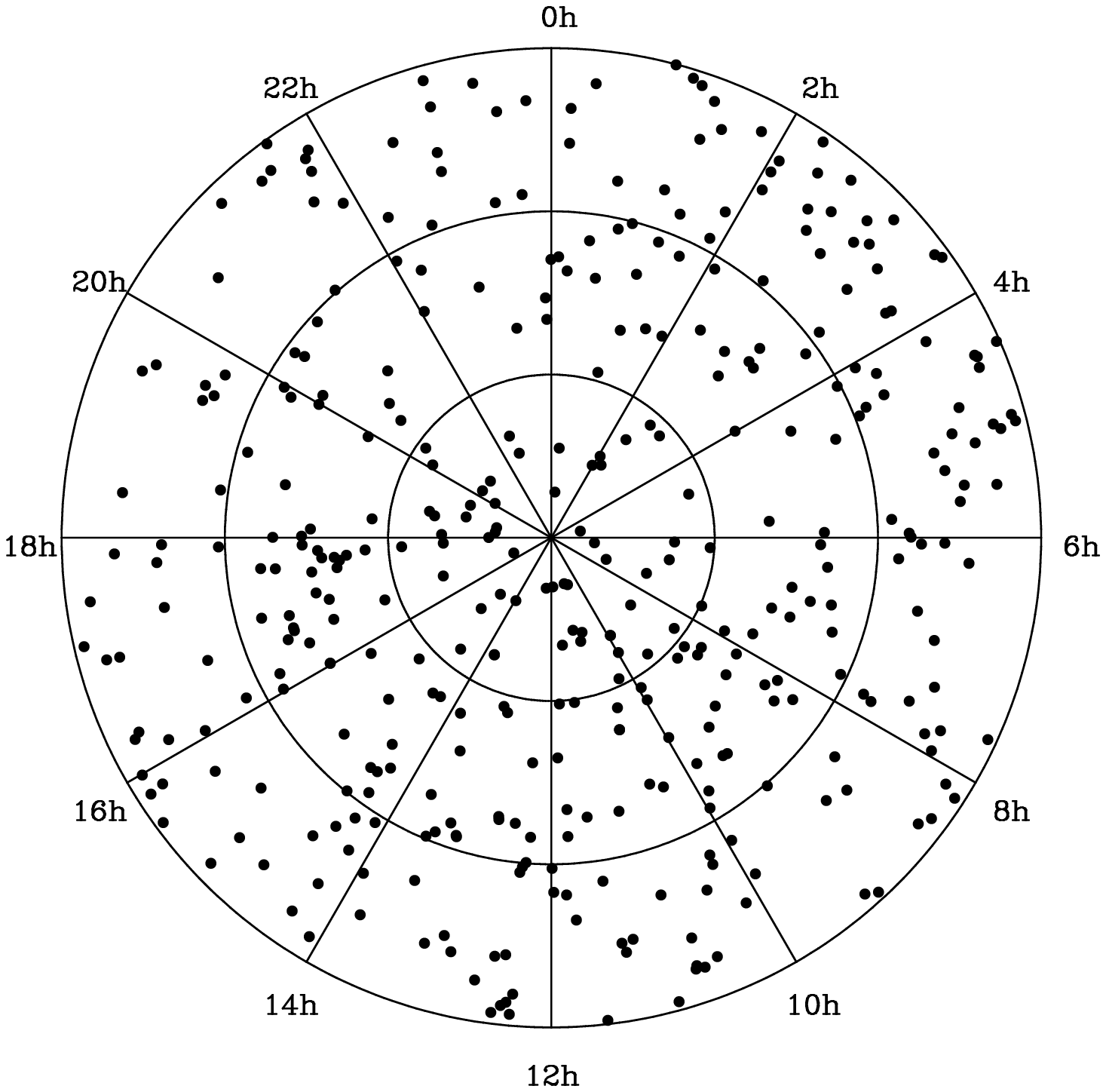}
\includegraphics[angle=0,origin=br,scale=0.50]{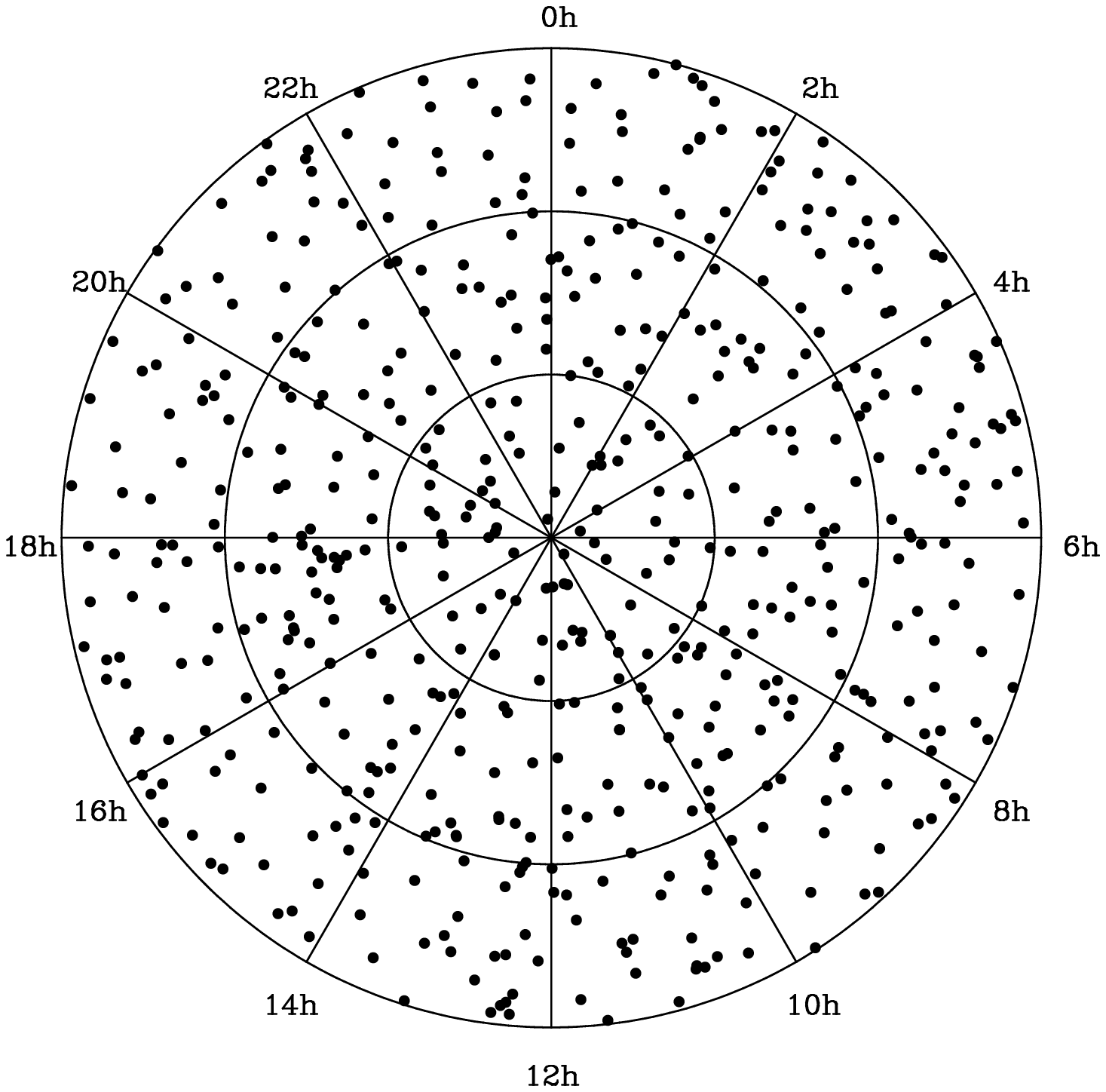}
\caption{\small Northern-sky source distribution in polar 
coordinates. {\it Left:} for the current ICRF, including defining, 
candidate, and ``other'' sources plus the additional sources 
published in ICRF-Ext.1 (see Fey et al.~\cite{Fey04}).
{\it Right:} same plot after adding the 150~new sources identified 
to fill the ``empty'' regions of the frame. The outer circle 
corresponds to a declination of~$0^{\circ}$ while the inner central 
point is for a declination of~$90^{\circ}$~. The intermediate circles 
correspond to declinations of $30^{\circ}$ and $60^{\circ}$.} 
\label{ICRF+150}
\end{figure*}

The approach used for selecting new sources to densify the ICRF
was to fill first the ``empty'' regions of the frame. The largest
such region for the northern sky is located near
$\alpha=$~22~h~05~min, $\delta=57^{\circ}$, where no ICRF source
is to be found within $13^{\circ}$. A new source should thus be
preferably added in that part of the sky. By using this approach
again and repeating  it many times, it is then possible to
progressively fill the ``empty'' regions of the frame and improve
the overall ICRF source distribution. The input catalog for
selecting the new sources to observe was the Jodrell Bank--VLA
Astrometric Survey (JVAS) which comprises a total 2118 compact
radio sources distributed over all the northern sky (Patnaik et
al. \cite{Pat92}, Browne et al. \cite{Bro98}, Wilkinson et al.
\cite{Wil98}). Each JVAS source has a peak flux density at 8.4~GHz
larger than 50~mJy at a resolution of 200~mas, contains $80\%$ or
more of the total source flux, and has a position known to
an rms accuracy of 12--55~mas. For every ``empty'' ICRF region, 
all JVAS sources within a radius of~$6^{\circ}$ (about 10~sources 
on average) were initially considered. These sources were then 
filtered out using the VLBA Calibrator Survey, which includes 
VLBI images at 8.4 and 2.3~GHz for most JVAS sources (Beasley 
et al. \cite{Bea02}), to eventually select the source with the 
most compact structure in each region.

The results of this iterative source selection scheme show that
30~new sources are required to reduce the angular distance to the 
nearest ICRF source from a maximum of~$13^{\circ}$ to a maximum 
of~$8^{\circ}$. Another 40~new sources would further reduce this 
distance to a maximum of~$7^{\circ}$ while for a maximum distance 
of~$6^{\circ}$, approximately 150~new sources should be added. 
Carrying this procedure further, it is found that the number of 
required new sources doubles for any further decrease of this 
distance of~$1^{\circ}$ (approximately 300~new sources for a maximum
distance of $5^{\circ}$ and 600~new sources for a maximum distance
of $4^{\circ}$) with the limitation that the JVAS catalog is not
uniform enough to fill all the regions below a distance of
$6^{\circ}$. Based on this analysis, we have selected the first
150~sources identified through this procedure for observation with
the EVN+ network described below. As shown in Fig.~\ref{ICRF+150},
the overall source distribution is potentially much improved with these
additional 150~sources in the northern sky.

\section{Observations and Data Analysis}

The observations were carried out in a standard geodetic mode during
three 24-hour dual-frequency (2.3 and 8.4~GHz) VLBI experiments
conducted on May~31, 2000, June~5, 2002, and October~27, 2003,
using the EVN (including the Chinese and South African telescopes)
and up to four additional geodetic radio telescopes (Algonquin 
Park in Canada, Goldstone/DSS~13 and Greenbank/NRAO20 in USA, 
and Ny-Alesund in Spitsbergen). There were between 10 and 12
telescopes scheduled for each experiment. Such a large network
permits a geometrically-strong schedule based on sub-netting which
allows tropospheric gradient effects to be estimated from the
data. The inclusion of large radio telescopes (Effelsberg,
Algonquin Park) in this network was essential because the new 
sources are much weaker than the ICRF ones (median total flux 
of 0.26~Jy compared to 0.83~Jy for the ICRF sources, see Charlot
et al.~\cite{Cha00}). Each experiment observed a total of 50~new 
sources along with 10~highly-accurate ICRF sources so that the 
positions of the new sources can be linked directly to the ICRF.

\begin{figure*}
\centering
\includegraphics[angle=-90,origin=br,scale=0.5]{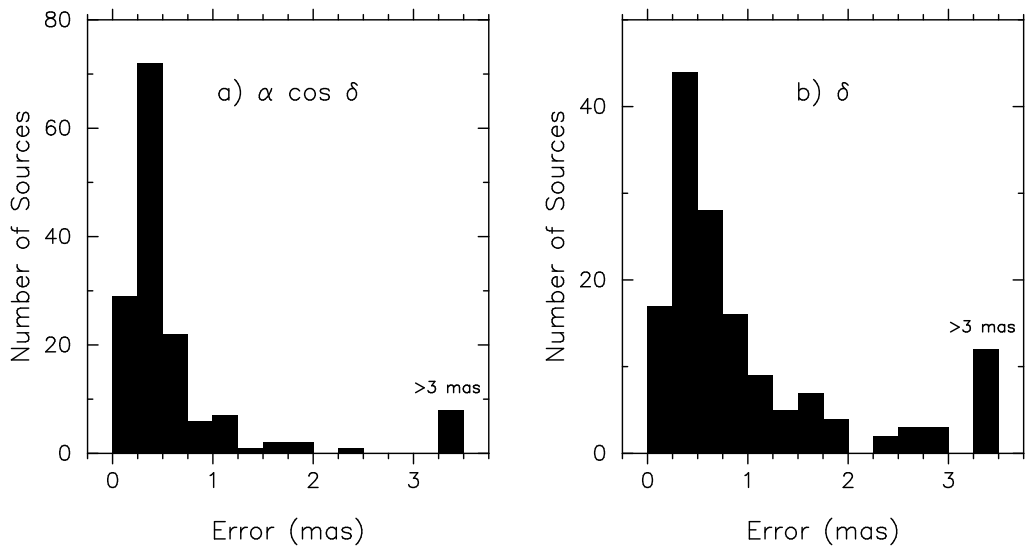}
\caption{\small Astrometric precision of the estimated coordinates
in {\it a)} right ascension and {\it b)} declination for the 
150~newly-observed sources. All errors larger than 3~mas are placed 
in a single bin marked with the label ``$>3$~mas'' on each plot.}
\label{Errors}
\end{figure*}

The data were correlated with the Bonn Mark~4 correlator,
fringe-fitted using the Haystack software fourfit, and exported in
the standard way to geodetic data base files. All subsequent
analysis employed the models implemented in the VLBI modeling and
analysis software MODEST (Sovers \& Jacobs \cite{Sov96}). Standard
geodetic VLBI parameters (station clock offsets and rates with
breaks when needed, zenith wet tropospheric delays every 3~hours,
and Earth orientation) were estimated in each experiment along
with the astrometric positions (right ascension and declination)
of the new sources. The positions of the 10~ICRF link sources 
were held fixed as were station coordinates. Observable weighting 
included added baseline-dependent noise adjusted for each baseline 
in each experiment in order to make $\chi^2$ per degree of freedom 
approximately equal to~1.

\section{Results}

The three EVN+ experiments described above have been very successful
in observing the selected targets. All 150~new potential ICRF sources have been
detected, hence indicating that the source selection strategy and
observing scheme set up for these experiments were appropriate. In 
the first two experiments (2000 May~31 and 2002 June~5), there were 
generally between 20 and 60~pairs of delay and delay rates usable 
for each source to estimate its astrometric position. Conversely, 
more than half of the sources observed in the third experiment 
(2003 October~27) had less than 20~pairs of usable delay and delay 
rates because of the failure of three telescopes in that experiment.

Figure~\ref{Errors} shows the error distribution in right ascension
and declination for the 150~newly-observed sources. The distribution 
indicates that about 70\% of the sources have position errors smaller 
than 1~mas, consistent with the high quality level of the ICRF. The 
median coordinate uncertainty is 0.37~mas in right ascension and 0.63~mas 
in declination. The larger declination errors are most probably caused 
by the predominantly East-West network used for these observations. 
Figure~\ref{Errors} also shows that a dozen sources have very large 
errors ($>3$~mas). Most of these sources were observed during the 
2003 October 27 experiment and have only a few available observations 
or data only on short intra-Europe baselines. Such sources should be 
re-observed to obtain improved coordinates if these are to be considered 
for inclusion in the next ICRF realization.

Among our 150 selected targets, 129 sources were found to have
astrometric positions available in the VLBA Calibrator Survey
(Beasley et al.~\cite{Bea02}). A comparison of these positions
with those estimated from our analysis shows agreement within
1~mas for half of the sources and within 2~mas for 80\% of the
sources. While the magnitude of the differences is consistent 
with the reported astrometric accuracy of the VLBA Calibrator
Survey, further investigation is necessary to determine whether 
these differences are of random nature or show systematic trends. 
Such trends may be caused by the limited geometry used in observing 
the VLBA Calibrator Survey (see Beasley et al.~\cite{Bea02}).

\section{Conclusion}

A total of 150 new potential ICRF sources have been successfully 
detected using the EVN and additional geodetic radio telescopes 
located in USA, Canada and Spitsbergen. About two-third of the 
sources observed with this EVN+ network have coordinate uncertainties 
better than 1~mas, and thus constitute valuable candidates for 
extending the ICRF. The inclusion of these sources would largely 
improve the ICRF sky distribution by naturally filling the ``empty'' 
regions of the current celestial frame.

Extending further the ICRF will require observing weaker and weaker 
sources as the celestial frame fills up and hence will depend closely 
on how fast the sensitivity of VLBI arrays improves in the future. 
Charlot~(\cite{Cha04}) estimates that an extragalactic VLBI celestial 
frame comprising 10~000 sources may be possible by 2010 considering 
foreseen improvements in recording data rates (disk-based recording, 
modern digital videoconverters) and new radio telescopes of the 
40--60~meter class that are being built, especially in Spain, Italy 
and China. In the even longer term, increasing the source density
beyond that order of magnitude is likely to require new instruments
such as the Square Kilometer Array envisioned by 2015--2020.

\begin{acknowledgements}
The European VLBI Network (EVN) is a joint facility of European, 
Chinese, South African and other radio astronomy institutes funded 
by their national research councils. The non-EVN radio telescopes 
in Algonquin Park (Canada), Ny-Alesund (Spitsbergen), Goldstone 
(USA), and Greenbank (USA) are sponsored by Natural Resources Canada, 
the Norvegian Mapping Authority, the National Aeronautics and Space 
Administration, and the U.~S. Naval Observatory, respectively. We 
thank all participating observatories, with special acknowledgements 
to the staff of the non-EVN geodetic stations for their enthousiasm 
in participating in this project. We are also grateful to Nancy 
Vandenberg for help in scheduling, Walter Alef and Arno Mueskens 
for data correlation and advice in fringe-fitting, and Axel Nothnagel 
for export of the data to geodetic data base files. This research 
was supported by the European Commission's I3 Programme ``RADIONET", 
under contract No.\ 505818.
\end{acknowledgements}

\end{document}